\documentclass[aps,prl,twocolumn,floats,superscriptaddress]{revtex4-2}
\usepackage{graphicx} 
\usepackage{amsmath}
\usepackage{amsfonts}
\usepackage{color}
\usepackage{xcolor}
\usepackage{braket}
\usepackage{verbatim}
\usepackage[colorlinks=true, linkcolor=blue, citecolor=blue, urlcolor=blue]{hyperref}

\newcommand{\be}{\begin{equation}}
\newcommand{\ee}{\end{equation}}
\newcommand{\bea}{\begin{eqnarray}}
\newcommand{\eea}{\end{eqnarray}}

\usepackage[normalem]{ulem}

\begin{document}

\title{Borromean Criticality in Two Dimensions}

\author{Alexandru Golic} 
\altaffiliation{These authors contributed equally to this work.}
\affiliation{Department of Physics, The Royal Institute of Technology, Stockholm SE-10691, Sweden}

\author{Igor Timoshuk} 
\altaffiliation{These authors contributed equally to this work.}
\affiliation{Department of Physics, The Royal Institute of Technology, Stockholm SE-10691, Sweden}
\affiliation{Wallenberg Initiative Materials Science for Sustainability, Department of Physics, The Royal Institute of Technology, Stockholm SE-10691, Sweden}

\author{Albert Samoilenka}
\affiliation{Clarendon Laboratory, University of Oxford, Oxford, UK}
\affiliation{Department of Physics, The Royal Institute of Technology, Stockholm SE-10691, Sweden}

\author{Egor Babaev} 
\affiliation{Department of Physics, The Royal Institute of Technology, Stockholm SE-10691, Sweden}
\affiliation{Wallenberg Initiative Materials Science for Sustainability, Department of Physics, The Royal Institute of Technology, Stockholm SE-10691, Sweden}

\author{Boris Svistunov} 
\affiliation{Department of Physics, University of Massachusetts, Amherst, MA 01003, USA}

\begin{abstract}

The characteristic feature of counterflow superfluids consisting of $N\geq 3$ components---the so-called Borromean supercounterfluids (BSCF)---is the presence of $N$ distinct elementary topological excitations (vortices) despite having only $N-1$ independent Goldstone modes. We show that this remarkable property clearly manifests itself at the Berezinskii-Kosterlitz-Thouless-type transition from the BSCF to the normal state, under the conditions of slight to moderate deviations from the case of exact intercomponent symmetry. 
More generally, our analysis also applies to any multicomponent superfluid with intercomponent drag fine-tuned to the value when certain composite vortices compete energetically with elementary ones.

\end{abstract}

\maketitle

{\it Introduction}---A peculiar family of ``super" states---the  counterflow superfluids, a.k.a. supercounterfluids (SCF) has been of substantial interest for more than two decades
\cite{Kuklov2003PhysRevLett.90.100401,
Babaev2002,
Lake22,
Altman2003,
Kuklov2004PhysRevLett.92.030403,
Kuklov2004PhysRevLett.92.050402,
Babaev2004,
Svistunov2015book,
Smiseth2005field,
Dahl2008PhysRevB.77.144519,
Hubener2009,
Powell2009,
Hu2009,
Soyler2009NJP11.073036,pollet2012recent,
Herland2010phase,
Menotti2010,
Hu2011,Duan2003,
Venegas_Gomez2020B,
deParny2021,
Basak2021,Venegas_Gomez2020A,
Blomquist2021PhysRevLett.127.255303,
Babaev2024PhysRevLett.133.026001,
Golic2025,
Kuklov2026PRB113.L060504,
Zheng2025,xiao2025fate}. 
The SCF regime takes place in a multicomponent (either bosonic or fermionic) system under the condition of arrested net superflow. At zero temperature, the SCF ground state is driven by Mott physics, requiring exact commensurability between the total particle density and the underlying lattice \cite{Kuklov2003PhysRevLett.90.100401}.   At finite temperature, SCF is driven by proliferation of a subset of multicomponent composite vortices
\cite{Babaev2002,Kuklov2004PhysRevLett.92.030403,Dahl2008PhysRevB.77.144519}. Another type of SCF state emerges when multiple fields of a multicomponent superconductor couple to a single U(1) gauge field, thereby suppressing the net matter flow in the system bulk 
through the Meissner effect but allowing for the neutral SCF modes \cite{Babaev2004} (see also Ref.~\cite{Svistunov2015book}).

Qualitatively new universal properties, distinguishing SCFs from the single-component superfluids, emerge in SCF systems with $N \geq 3$ components---the so-called Borromean SCF (BSCF) \cite{Svistunov2015book,
Blomquist2021PhysRevLett.127.255303,
Babaev2024PhysRevLett.133.026001,
Golic2025,
Kuklov2026PRB113.L060504}.
The most prominent feature of BSCF 
is the presence of $N$ distinct elementary topological excitations (vortices) despite having only $N-1$ independent Goldstone modes.  BSCF is a parent state for a novel time-reversal-breaking ``Borromean insulator" phase driven by a frustrating intercomponent 
Josephson coupling \cite{Bojesen2013,Bojesen2014, Babaev2024PhysRevLett.133.026001,Grinenko2021,Shipulin2023}.

A natural fundamental question is of the universality class of the temperature-driven BSCF--to--normal-fluid (BSCF--NF) phase transition. 
It was argued \cite{Kuklov2026PRB113.L060504} that in three dimensions (3D), the transition is inevitably of the first order; the outcome 
of the simulation of the three-component Bose-Hubbard model \cite{Blomquist2021PhysRevLett.127.255303} is consistent with this prediction. The situation in two dimensions (2D) is radically different. Here, the BSCF--NF transition is supposed to be of the Berezinski-Kosterlitz-Thouless (BKT) type---driven by dissociation of the vortex-antivortex pairs; therefore, it is natural to expect that the presence of $N$ types of distinctively different interacting pairs might have a pronounced qualitative effect on the criticality. A support for such an expectation comes from the fact that the very nature of the BSCF state described by the compact-gauge-invariant effective theory implies strong drag between different components \cite{Babaev2024PhysRevLett.133.026001,Golic2025}, resulting in a nontrivial interplay between vortex-antivortex pairs of different flavors.    

For BSCF with exact intercomponent symmetry, the theory of BKT transition was developed in Ref.~\cite{Golic2025}. Counterintuitively enough, the resulting renormalization-group (RG) equations turn out to be the same as in the original theory by Kosterlitz and Thouless \cite{KT} because (i) all the $N$ superfluid stifnesses and all the $N$ densities of vortex-antivortex pairs are identical reducing the number of the RG equations to just two and (ii) the number of types of pairs and the density of pairs enter the theory in the form of a global factor thus excluding the possibility of observing this number by tracing the flow of the superfluid stiffness.

In this Letter, we show how a slight-to-moderate asymmetry between the components changes the structure of the RG equations, thereby making the number of distinct vortex-antivortex pairs a crucial parameter that defines both the quantitative and qualitative aspects of the RG flow. Of special importance for the resulting theory is the fact---first observed in 2D in Ref.~\cite{Karle2019} (and in 3D in Ref.~\cite{Dahl2008PhysRevB.77.144519}) that the drag between components in a multicomponent superfluid preserves the single transition from splitting into a sequence of single-component transitions as long as asymmetry between the components is appropriately small.

\begin{figure}
   \centering
\includegraphics[width=1.00\linewidth]{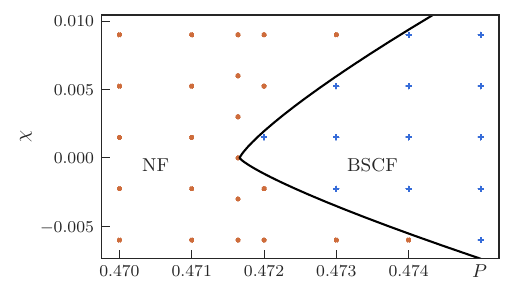}
    \caption{Phase diagram of the J-current model in the close vicinity of the critical point of the component-symmetric regime taking place at $P=P_0=0.4716$, $\chi=0$.  The phase boundaries have the shape $\chi \propto \pm (P-P)^{\varphi/2}$ controlled by the golden-ratio exponent $\varphi$; this shape comes from Eqs.~(\ref{phase_boudaries})--(\ref{xi_c}) with parameters  $C$ and $D$ being linear functions of $P$ and $\chi$, respectively. The dots indicate the model parameters for which numerical simulations were performed, and the results demonstrate a perfect match with RG predictions. Red (blue) dots correspond to the normal (BSCF) state. At each point we studied 12 different sample sizes, from $L=40$ to $L=300$ (see, e.g., Fig.~\ref{fig:w_s_dep}), performing $\sim 5 \cdot 10^{13}$ Monte Carlo steps upon thermalization.}
    \label{fig:RD_phase_diag}
\end{figure}

Specifically, we analyze a two-component representation of the three-component BSCF, Eq.~(\ref{Action_density_gen}). The qualitative difference with the two-component setup of  Ref.~\cite{Karle2019} is that here the strength of the drag between the two components is fine-tuned in such a way that composite vortices become energetically similar to the single-component ones, thus realizing the situation where there are more vortex species than Goldstone modes. In the BSCF context, these composite vortices represent the elementary vortices of the third component. However, the model has a broader context that we refer to as {\it enforced Borromeanicity}. 
A signature feature of the theory is the peculiar power-law scaling of the asymmetry effects---including the shape of the phase boundaries, see Fig.~\ref{fig:RD_phase_diag}---involving the golden ratio constant $\varphi=\left(1+\sqrt{5}\right)/2$.

{\it Effective action}---Consider a three-component BSCF featuring exact intercomponent symmetry. The long-wave properties of the system are described by the following compact-gauge-invariant action density \cite{Golic2025}  
\be
{\cal A}[\{ \theta \}]  = \frac{\Lambda_s}{4} \sum_{\alpha < \beta} (\nabla \theta_\alpha - \nabla \theta_\beta)^2 \, ,
\label{Action_density_sym}
\ee
where $\theta_\alpha$ ($\alpha=1,2,3$) are the phases of the three components.
Rewriting (\ref{Action_density_sym}) in terms of two relative phases, $\phi_{1,2} = \theta_{1,2} - \theta_3$ (or, equivalently, fixing the gauge by requiring $\theta_3 \equiv 0$), we get the action density
\begin{align}
    \mathcal{A}[\{\phi\}] = \frac{\Lambda_s}{2}\left[\left(\nabla\phi_1\right)^2 + \left(\nabla\phi_2\right)^2\right] - \frac{\Gamma}{2}\, \nabla\phi_1\cdot\nabla\phi_2 \, ,
    \label{Action_density_gen}
\end{align}
with $\Gamma = \Lambda_s$. In a broader context, action (\ref{Action_density_gen}) with $\Gamma$ not necessarily equal to $\Lambda_s$ can be interpreted as a two-component superfluid with Andreev--Bashkin drag---the coupling between the currents of the two components. A two-component system with a generic drag has only two types of elementary vortices. A very special (exotic) regime takes place when the drag coefficient  $\Gamma$ is fine-tuned to be close (equal) to $\Lambda_s$, in which case the composite topological excitation consisting of the elementary vortices in each of the two components has the energy close (equal) to the energy of the elementary single-component vortex. In the original Borromean notation, this composite vortex corresponds to the elementary vortex in component 3. That is why we refer to such a regime in a fine-tuned two-component system as the {\it enforced Borromeanicity} regime. 

Along similar lines, observe that the simple transformation $\phi_1 \to - \phi_1$ maps the model (\ref{Action_density_gen}) onto itself with an opposite sign of $\Gamma$.  Hence, the BKT physics of the model proves insensitive to the sign of $\Gamma$---up to changing the sense of rotation of the vortices of one of the two components---and should be described by the same set of RG equations. 

{\it RG equations}---At the symmetric point $\Gamma = \Lambda_s$ the two types of single-component vortex pairs have the same energy as the composite (1, 1) vortex pair, and the RG equations are identical to the symmetric Borromean case discussed in Ref.~\cite{Golic2025}.
If the drag is shifted away from this point, then the energy of a composite vortex pair $\Lambda_d = 2\Lambda_s - \Gamma$ will correspondingly differ from the single-component ones.
This asymmetry leads to a different RG flow for the single- and composite vortices, resulting in the following system of four RG equations: 
\bea
    \frac{\partial}{\partial\lambda}\Lambda_s &=& -\Lambda_s^2g_s -\frac{1}{4}(2\Lambda_s - \Lambda_d)^2g_s - \frac{1}{4}\Lambda_d^2g_d\, , \\
    \frac{\partial}{\partial\lambda}\Lambda_d &=&  -\Lambda_d^2g_d -\frac{1}{2}\Lambda_s^2g_d\, , \\
    \frac{\partial}{\partial\lambda}g_s &=&  -2(\pi\Lambda_s - 2)g_s\, , \quad
    \frac{\partial}{\partial\lambda}g_d \, =\,   -2(\pi\Lambda_d - 2)g_d \, , \qquad
\eea
where $\lambda = \ln (l/l_0)$ is the logarithm of the length scale $l$ measured in units of the characteristic microscopic scale $l_0$; $g_s$ and $ g_d$ are the fugacities of the single- and composite vortex pairs, respectively, at the separation order $l$.

Being interested in a close vicinity of the component-symmetric transition point, we expand RG equations up to the leading order in powers of  $w_{s,d} = \pi\Lambda_{s, d} - 2 \ll 1$:
\begin{align}
    &w_s' = -5g_s - g_d, \label{RG1}\\
    &w_d' = -4g_d - 2g_s, \label{RG2}\\
    &g_s' = -2w_sg_s, \label{RG3}\\
    &g_d' = -2w_dg_d. \label{RG4}
\end{align}
It is instructive to compare the system of RG equations (\ref{RG1})--(\ref{RG4}) with that of Ref.~\cite{Karle2019}. An immediate observation is that despite having one more type of vortex pairs, we have fewer equations: four as opposed to five in Ref.~\cite{Karle2019}. An obvious reason for the difference is that---for the sake of simplicity---we are not addressing the general asymmetric case when the stiffnesses of components 1 and 2 are different. In the latter case, we would have six equations: three describing the flow of three different stiffnesses and three describing the flow of the fugacities of three different vortex-antivortex pairs. What the two RG systems have in common is that the criticality is associated with reaching the universal Nelson-Kosterlitz (NK) condition by (at least) one of the vortex-antivortex subsystems. In our notation, this corresponds to $w_s(\infty)=0$ or $w_d(\infty)=0$.

{\it Constant of RG flow}---From the structure of right-hand sides of the RG equations it is clear that there exists a constant of the flow, ${\cal S} \equiv {\cal S}(w_s, w_d,g_s, g_d )$, having a relatively simple form---quadratic in the variables $(w_s, w_d)$ and linear in the variables $(g_s, g_d)$. Corresponding substitution into the RG equations readily fixes the coefficients of the form (up to a global factor):
\be
{\cal S}\, =\, 8w_s^2 + 5w_d^2 - 4 w_s w_d - 36 g_s - 18g_d \, .
\label{cal_S}
\ee
In the superfluid phase, $g_s(\infty)=g_d(\infty)=0$ and the constant ${\cal S}$ thus provides the relation between the thermodynamic values of $w_s$ and $w_d$:
\be
8w_s^2(\infty) + 5w_d^2(\infty) - 4 w_s(\infty) w_d(\infty) \, =\, {\cal S} \, .
\label{S_asympt}
\ee
This relation is particularly important at the critical point where one of the two components reaches the universal NK value. Here the constant ${\cal S}$ controls the amplitude of the critical value (and thus the value of nonuniversal jump across the transition point) of the stiffness of the other component: $w_s^{(c)}(\infty) = \sqrt{{\cal S}/8}$,  $w_d^{(c)}(\infty) = \sqrt{{\cal S}/5}$.
At the symmetric critical point, we have $w_s(\infty)=w_d(\infty)=0$ meaning ${\cal S}=0$.

{\it UV asymptotics}---At the symmetric Borromean  critical point, the functions  $w_s$, $w_d$, $g_s$, and $g_d$ are known to have the following analytical form: $w_{s,d}(\lambda)=1/\lambda$, $g_{s,d}(\lambda)=1/6\lambda^2$ \cite{Golic2025}.
In the vicinity of this point,  calculate the first-order deviations from this critical solution by solving the linearized system of flow equations; the result is 
\begin{align}
    &w_s = \frac{1}{\lambda} +\frac{C}{3}\lambda + \frac{C'}{\lambda^2} - \frac{\varphi D}{2}\lambda^{1/\varphi} + \frac{D'}{2\varphi}\lambda^{-\varphi}, \label{sol1} \\
    &w_d = \frac{1}{\lambda} +\frac{C}{3}\lambda + \frac{C'}{\lambda^2} + \varphi D\lambda^{1/\varphi} - \frac{D'}{\varphi}\lambda^{-\varphi}, \label{sol2} \\
    &g_s = \frac{1}{6\lambda^2} - \frac{C}{18} + \frac{C'}{3\lambda^3} + \frac{D}{6}\lambda^{1/\varphi - 1} + \frac{D'}{6}\lambda^{-\varphi - 1}, \label{sol3}\\
    &g_d = \frac{1}{6\lambda^2} - \frac{C}{18} + \frac{C'}{3\lambda^3} - \frac{D}{3}\lambda^{1/\varphi - 1} - \frac{D'}{3}\lambda^{-\varphi - 1}, \label{sol4}
\end{align}
with $\varphi=\left(1+\sqrt{5}\right)/2$ (the golden ratio constant). Below, we argue that the constants $C'$ and $D'$ should be set to zero, thus leaving us with a two-parameter solution controlled by the constants $C$ and $D$. 

The meaning of the constant $C'$ is very transparent. It is directly related to the exact invariance of RG equations (\ref{RG1})--(\ref{RG4}) with respect to the distance-rescaling transformation $\lambda\rightarrow\lambda+\lambda_0$.  Hence, without loss of generality, we absorb the constant $C'$ into the definition $\lambda = \ln (l/l_0)$ that naturally features a freedom of choosing the scale of distance $l_0$.

Turning to the constant $D'$, our first observation is that this constant is irrelevant in the long-wave limit, where it generates a subleading correction only. Our next observation is that in the opposite limit of $\lambda \to 0$, the corresponding term should also remain subleading. Indeed, from the solution (\ref{sol1})--(\ref{sol4}) we see that nonzero value of $D'$ characterizes asymmetry between $s$- and $d$-functions. Since all our analysis is based on the assumption that we are very close to the symmetric regime, we conclude that setting $D'=0$ is appropriate for any relevant value of $\lambda$. In contrast, the constant $D$, which is also associated with asymmetry and thus has to be small, generates a perturbation that becomes relevant in the $\lambda \to \infty$ limit, provided the constant $C$ is sufficiently small.
 
Hence, constants $C$ and $D$ fully control the shape of all four functions at any $\lambda$, from $\lambda \to +0$ to $\lambda \to +\infty$. In particular, the constant ${\cal S}$ can be expressed in terms of them by substituting (\ref{sol1})--(\ref{sol4}) into (\ref{cal_S}). This way, we find that ${\cal S}$ is independent of $D$ and, up to a numerical coefficient, equals $C$: ${\cal S} = 9C$.

\begin{figure}
    \centering
    \includegraphics[width=\linewidth]{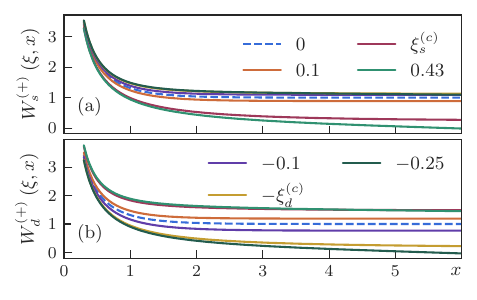}
    \caption{Scaling functions $W_{s}^{(+)} (\xi, x )$ and $W_{d}^{(+)} (\xi, x )$  defined in Eqs.~(\ref{scaling_functions}) and (\ref{xi}). Different colors correspond to different values of parameter $\xi$, as indicated in the legends.}
    \label{fig:W_plot}
\end{figure}

{\it Scaling}---The system of RG equations (\ref{RG1})--(\ref{RG4}) is invariant with respect to the following scaling transformation
\be
\lambda \, \to \, \zeta \lambda \, , \quad  w_{s,d} \, \to \, w_{s,d}/\zeta \, , \quad
 g_{s,d} \, \to \, g_{s,d}/\zeta^2  ,
\label{scaling}
\ee
meaning that (\ref{scaling}) is supposed to convert any solution of the system (\ref{RG1})--(\ref{RG4}) into yet another solution with different free constants, whose rescaling matches (\ref{scaling}). With the asymptotic expressions (\ref{sol1})--(\ref{sol4}) we see that the corresponding transformation is
\be
C \, \to \, C/\zeta^2 \, ,  \qquad D \, \to \, D/\zeta^\varphi \, .
\label{match}
\ee
This brings us to the scaling form of the solutions (the superscript corresponds to the sign of the constant $C$):
\bea
w_{s,d} &=& \,\sqrt{|C|} \, W_{s,d}^{(\pm)}(\xi,\,  x) , \quad 
g_{s,d} \, =\,   |C| \, G_{s,d}^{(\pm)}(\xi, \, x), \qquad
\label{scaling_functions} \\
x &=& \sqrt{|C|} \lambda \, , \qquad
\xi \, =\,  D/|C|^{\varphi/2} \, ,
\label{xi}
\eea
see Fig.~\ref{fig:W_plot} for an illustration;
Note an important role played by the scaling parameter $\xi$. In the  $x\to 0$ limit, the leading terms and subleading corrections of these functions are dictated by Eqs.~(\ref{sol1})--(\ref{sol4}) with $C'=D'=0$. The ``$-$" functions are related to their ``$+$" counterparts by the requirement that at any finite $\lambda$ the quantities  $w_{s,d}$ and $g_{s,d}$ be analytic functions of $C$ and $D$.

\begin{figure}
    \centering
    \includegraphics[width=\linewidth]{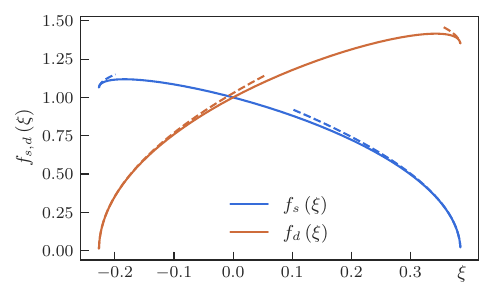}
    \caption{Numerically calculated functions $f_s\left(\xi\right)$ and $f_d\left(\xi\right)$ defined within the interval $\xi \in [-\xi_d^{(c)} , \xi_s^{(c)} ]$. The dashed lines are the fits to corresponding asymptotic regimes (\ref{f_s_crit})--(\ref{f_s_crit2}).}
    \label{fig:f_full}
\end{figure}

At a  $C>0$ and small enough  $\xi$, the scaling functions are supposed to have finite limits at $x \to +\infty$ (meaning that RG flow results in the supercounterfluid state):
\bea
G_{s,d}(\xi, +\infty) &=& 0 \qquad  (|\xi| \leq \xi^{(c)}_{s,d}) \label{limits1} \\
W_{s,d}(\xi, +\infty) &=& f_{s,d}(\xi) \, \geq \, 0 \,  \qquad (|\xi| \leq \xi^{(c)}_{s,d}) \, .\qquad
\label{limits2}
\eea
Here $\xi^{(c)}_{s,d}$ are two critical values of $\xi$ such that
\bea
f_s\left(\xi^{(c)}_s\right) &=&  0\, , \qquad f_d\left(-\xi^{(c)}_d\right) \, =\, 0\, , 
\label{crit_cond} \\
W_s\left(\xi > \xi^{(c)}_s  , \, +\infty \right) &=&  
W_d \left(\xi < -\xi^{(c)}_d  , \,  +\infty \right) \, =\,  -\infty \, . \nonumber
\eea
The functions $f_{s,d}(\xi)$ and thus the values of $\xi^{(c)}_{s,d}$ can be found by numerically integrating the RG equations [see Fig.~\ref{fig:f_full}],   yielding, in particular, precise relations for the two phase boundaries
\bea
D &=& \xi^{(c)}_s C^{\varphi/2}\, , \quad D\, =\,  - \xi^{(c)}_d C^{\varphi/2} \qquad (C \geq 0) \, , \quad
\label{phase_boudaries} \\
\xi_s^{(c)} &\approx& 0.385 \, , \qquad   \xi_d^{(c)} \approx 0.227 \, .
\label{xi_c} 
\eea
By their definitions, the scaling functions $f_{s,d}(\xi)$ fully define the dependence of the thermodynamic values $w_{s,d}(\infty)$ on the parameters $C$ and $D$: $w_{s,d}(\infty) \, =\, \sqrt{C} f_{s,d}(D/C^{\varphi/2})$.
Substituting this into (\ref{S_asympt}) and taking into account that ${\cal S} = 9C$, we obtain
\be
8f_s^2(\xi) + 5f_d^2(\xi) - 4 f_s(\xi)f_d(\xi) \, =\, 9 \, .
\label{f_s_d_relation}
\ee
On the approach to the critical points, $\xi \to \xi_s^{(c)} - 0$ or $\xi \to  -\xi_d^{(c)}$+0, corresponding functions $f_{s,d}$ demonstrate standard BKT behavior (see End Matter):
\bea
f_s(\xi) &\to&  A_s \sqrt{\xi_s^{(c)} - \xi} \quad \text{at} \quad \xi \, \to  \,\xi_s^{(c)} - 0 \, ,\quad 
\label{f_s_crit} \\
f_d(\xi) &\to& A_d \sqrt{\xi_d^{(c)} + \xi} \quad \text{at} \quad \xi \, \to  \,- \xi_d^{(c)} +0 \, , \quad
\label{f_d_crit} \\
A_s &\approx& 1.73 \, , \qquad A_d \, \approx\, 2.16 \, .
\eea
Substituting (\ref{f_s_crit}) and (\ref{f_d_crit}) into (\ref{f_s_d_relation}) and keeping only the leading (constant) and subleading (the square-root) terms, we obtain two counterparts of (\ref{f_s_crit}) and (\ref{f_d_crit}):
\bea
f_d(\xi) &\to& \frac{3}{\sqrt{5}} +  \frac{2 A_s}{5} \sqrt{\xi_s^{(c)} - \xi} \quad \text{at} \quad \xi \, \to  \,\xi_s^{(c)}-0 ,\qquad
\label{f_d_crit2} \\
f_s(\xi) &\to& \frac{3}{\sqrt{8}} +  \frac{A_d}{4} \sqrt{\xi_d^{(c)} + \xi} \quad \text{at} \quad \xi \, \to  - \xi_d^{(c)}+0 .\qquad
\label{f_s_crit2}
\eea
{\it Numeric simulations}---We validated our theory against simulations of the minimal Borromean loop (a.k.a. J-current) model introduced in Ref.~\cite{Golic2025}. The configurational space of the model consists of three kinds of oriented loops living on the bonds of a square lattice. The Borromeanicity is hard-wired by the requirement that the state of each bond, $(n_1,n_2,n_3)$, is such that the algebraic sum of all three bond currents is zero: $n_1+n_2 + n_3=0$. The model's minimalism requires that the magnitudes of the three integers be either 0 or 1. (In the loop language, this means that the loops of the same color, while being allowed to intersect at some sites, cannot overlap at any bond.)
Hence, there are only four different bond states (up to the flipping of the directions of all the bond currents): $(0,0,0)$, $(\pm 1, \mp 1, 0)$, $(\pm 1, 0, \mp 1)$, and $(0, \pm 1, \mp 1)$. The symmetry between components 1 and 2 implies that the bond states $(\pm 1, 0, \mp 1)$ and $(0, \pm 1, \mp 1)$ have the same weight, $s_w$. The weight $c_w$ of the bond $(\pm 1, \mp 1, 0)$ can be different from $s_w$, thus allowing component 3 to be different from components 1 and 2.
Following the numeric protocol of Ref.~\cite{Golic2025}, we sample configurations with the worm algorithm \cite{Prokofev1998a,Prokofev1998b,Prokofev2001} and estimate the superfluid stiffnesses from the statistics of worldline winding numbers.

\begin{figure}
    \centering
    \includegraphics[width=\linewidth]{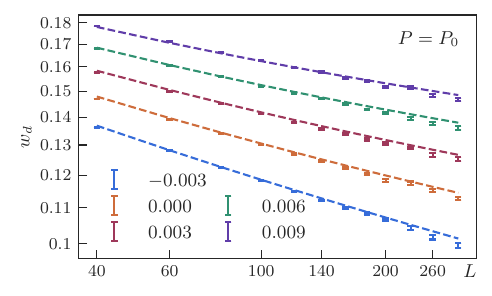}
    \caption{The flow of $w_d$ with the system size at $P=P_0$ and various values of $\chi$ indicated in the legend. The data points with error bars are obtained by Monte Carlo simulation of the J-current model. The dashed lines are the RG predictions. The only fitting parameter is the coefficient $D_0$ in the relation $D=D_0 \chi$. The values of the other parameters are already fixed by the properties of the symmetric regime $\chi = 0$.}
    \label{fig:w_s_dep}
\end{figure}

Our RG treatment applies to the close vicinity of the normal-to-BSCF transition of the symmetric model. In that region, the parameters $C$ and $D$ are related to $s_w$ and $c_w$ by a linear transformation. Since both $D$ and $\chi=s_w-c_w$ characterize the asymmetry of the model, they have to be proportional to each other: $D=D_0 \chi$. Parameter $C$ characterizing the symmetric properties of the system has to be a linear {\it symmetric} function of the three weights of the three different non-empty bonds, that is, $C$ should be a linear function of $s_w+s_w+c_w = 2s_w+c_w$. Hence,  $C=C_0\left(P-P_0\right)$, where $P=\left(2s_w+c_w\right)/3$; with $C_0=2.61$ and $P_0=0.4716$ being the same as in the symmetric model considered in Ref.~\cite{Golic2025}. Another important parameter, already known from the simulation of the symmetric case, is the length $l_0$, which enters the definition $\lambda=\ln(L/l_0)$. The only new parameter is $D_0=0.678$, the value of which we fix by fitting the simulation results with solutions of RG equations. In this way, we obtain the phase diagram shown in Fig.~\ref{fig:RD_phase_diag}. The figure also shows a set of points in both the normal and BSCF phases at which the J-current model was simulated to ensure that the RG equations accurately capture the flow of superfluid stiffnesses as the system size increases. A characteristic example of corresponding comparison is presented in Fig.~\ref{fig:w_s_dep}.

{\it Conclusions}---In an instructive contrast to the higher-dimensional cases, where the temperature-driven phase transition from the state of Borromean supercounterfluid (BSCF) to the normal state is either of the first order or splits into componentwise transitions, the two-dimensional case is characterized by a direct, BKT-type transition. In the component-asymmetric case, the transition does not split into componentwise transitions as long as the asymmetry is appropriately small. Furthermore, having a slight asymmetry is crucial for revealing the distinct BSCF-specific features of the RG flow.

The simplest case that captures the essence of the difference between the BSCF-to-normal transition and the standard BKT transition is the three-component BSCF with one component somewhat different from the other two. The compact-gauge-invariant three-component action of this model maps onto a two-component action (\ref{Action_density_gen}) featuring exact symmetry between the components 1 and 2, with the third component being implicitly represented by the fine-tuned drag coefficient.
Action (\ref{Action_density_gen}) acquires a broader context if literally understood as a two-component model fine-tuned to the regime $\Gamma \approx \Lambda_s$ that we refer to as enforced Borromeanicity. The regime of enforced Borromeanicity is isomorphic (via the transformation $\phi_1 \to - \phi_1$, which changes the sign of $\Gamma$) to the regime when single-component ordering competes with the ordering in the pairing channel.
{

The RG flow is described by four coupled equations (\ref{RG1})--(\ref{RG4}) in terms of two superfluid stiffnesses and two fugacities of vortex-antivortex antivortex pairs. The signature feature of the flow is the scaling symmetry (\ref{scaling})--(\ref{match})---curiously involving the golden-ratio exponent $\varphi$---implying the scaling form (\ref{scaling_functions})--(\ref{xi}) [see also Fig.~\ref{fig:W_plot}] of the solutions and, in particular, Eqs.~(\ref{phase_boudaries})--(\ref{xi_c}) for the phase boundaries, as well as Eqs.~(\ref{f_s_crit})--(\ref{f_s_crit2}) [see also Fig.~\ref{fig:f_full}] for the critical behavior of superfluid stiffnesses.

We validated our theory against a Borromean J-current model simulated with the worm algorithm. We observed a perfect match of the flows of superfluid stiffnesses with that predicted by the RG equations (see Fig.~\ref{fig:w_s_dep}), which, in particular, allowed us to obtain accurate results for the phase diagram of the J-current model in the vicinity of the component-symmetric critical point,  Fig.~\ref{fig:RD_phase_diag}.

In general, the number of distinct elementary topological excitations can be much larger than the number of Goldstone modes. We demonstrate this in the End Matter section by showing that the maximal number of elementary vortices grows as half the lattice kissing number in space with dimension equal to the number of Goldstone modes.

{\it Acknowledgments}---
This work was supported by the Swedish Research Council Grant 2022-04763, by the Knut and Alice Wallenberg Foundation Project No. KAW 2024.0131 and the Wallenberg Initiative Materials Science for Sustainability (WISE) funded by the Knut and Alice Wallenberg Foundation. BS acknowledges support from the National Science Foundation under Grant DMR-2335904.
AS acknowledges support from the Swedish Research Council (VR) through Grant No. 2025-07901.
The simulations were enabled by resources provided by the National Academic Infrastructure for Supercomputing in Sweden (NAISS), partially funded by the Swedish Research Council through grant agreement no. 2022-06725.

\bibliography{enbor}

\newpage

\begin{widetext}
\section{End Matter}
\end{widetext}

{\it Details of RG flow} depend on the sign of $C$ and the value of the parameter $\xi$ (\ref{xi}).
There are five characteristic regimes. 

\noindent {\it (i) Case $|\xi| \ll 1$.} Here, the role of asymmetry reduces to subleading corrections that can be safely ignored.

\noindent {\it (ii) Case $C>0$  with $-\xi_d^{(c)} < \xi < \xi_s^{(c)}$ and $|\xi - \xi_s^{(c)}| \sim |\xi+ \xi_d^{(c)}|  \sim 1$.} At the qualitative level, this case is very reminiscent of the symmetric superfluid case taking place at $C>0$.
In particular, the logarithm of correlation length, $\lambda_c$, is defined by the same condition $\lambda_c \sim 1/\sqrt{C}$. At $\lambda \ll \lambda_c$, the flow is quantitatively the same---up to subleading corrections, as in the symmetric case. Quantitative difference with the symmetric case takes place at $\lambda \sim \lambda_c$, at which scale the flow saturates to the thermodynamic limit.

\noindent {\it (iii) Case $C>0$ with $|\xi - \xi_s^{(c)}| \ll 1$.} Here we are in the vicinity of the critical line $\xi = \xi_s^{(c)}$. The flow is characterized by two correlation lengths. The lower one is the above-discussed $\lambda_c \sim 1/\sqrt{C}$. As opposed to cases (i) and (ii), the flow does not yet saturate at $\lambda_c \sim 1/\sqrt{C}$. Rather, it changes its character. Since $w_d$ stays larger than $\sqrt{9C/5}$---see Eq.~(\ref{f_d_crit2})---the fugacity $g_d$ gets exponentially suppressed at $\lambda\gg \lambda_c$ and can be set equal to zero. As a result, Eqs.~(\ref{RG1}) and (\ref{RG3}) decouple into the closed system 
\begin{align}
    &w_s' = -5g_s \, , \label{RG1_s}\\
    &g_s' = -2w_sg_s\, , \label{RG3_s}
\end{align}
Eq.~(\ref{RG4}) is trivially satisfied, and (\ref{RG2}) combined with (\ref{RG1_s}) yields
\be
w_d' \, =\, \frac{2}{5}\, w_s' \quad \Rightarrow \quad w_d \, =\, \frac{2}{5}\, w_s \, + \, 3\, \sqrt{C/ 5} \, .
\label{w_d}
\ee
The constant in the second equality is fixed by the requirement of matching (\ref{f_d_crit2}) up to higher-order corrections. 

Equations (\ref{RG1_s}) and (\ref{RG3_s}) have the standard Kosterlitz-Thouless form and thus are immediately solved, yielding
\bea
w_s(\lambda) &=& \sqrt{Q_s} \coth \left( \sqrt{Q_s} \lambda \right)  \qquad (Q_s > 0)\, , \\
w_s(\lambda) &=& \sqrt{|Q_s|} \cot \left( \sqrt{|Q_s|} \lambda \right)  \qquad (Q_s < 0)\, .
\eea
The constant $Q_s$ needs to be related to the constants $C$ and $D$. Given that $Q_s$ vanishes at the critical point, and we deal with a closed vicinity of the critical line, we can confine ourselves to linear in $\xi$ terms. This brings us to the expression 
\be
Q_s \, = \, C A_s^2 \, (\xi_s^{(c)} -\xi) \, ,
\ee
where the proportionality coefficient is fixed by the requirement that it matches Eq.~(\ref{f_s_crit}). Since we are dealing with the regime $|\xi - \xi_s^{(c)}| \ll 1$, the magnitude of $Q_s$ is much smaller than $C$, thus defining (the logarithm of) the upper correlation radius: $\lambda_c^{\rm (upper)} \sim 1/\sqrt{|Q_s|} \gg \lambda_c$.

\noindent {\it (iv) Case $C>0$ with $|\xi + \xi_d^{(c)}| \ll 1$.}
This case is directly analogous to the previous one, except that the subscripts $s$ and $d$ are swapped and some numerical coefficients are changed. We thus write down the final results:

\bea
w_d(\lambda) &=& \sqrt{Q_d} \coth \left( \sqrt{Q_d} \lambda \right)  \qquad (Q_d > 0)\, , \\
w_d(\lambda) &=& \sqrt{|Q_d|} \cot \left( \sqrt{|Q_d|} \lambda \right)  \qquad (Q_d < 0)\, ,~~\\
Q_d &=& C A_d^2 \, (\xi_d^{(c)} + \xi) \, ,~~
\\
 w_s &=& \frac{1}{4}\, w_d \, +\,  3\, \sqrt{C/8} \, .
\label{w_s}
\eea
For the logarithm of the upper correlation radius, we have $\lambda_c^{\rm (upper)} \sim 1/\sqrt{|Q_d|} \gg \lambda_c$.

\noindent {\it (v) Cases $\xi - \xi^{(c)}_s \gtrsim 1$ and $- \xi -\xi^{(c)}_d \gtrsim 1$.} These two very similar cases correspond to the normal state. What distinguishes them from case (i) with negative $C$ is that the asymmetry now plays an important role. As is clear from Eqs.~(\ref{sol1})--(\ref{sol4}) it is the parameter $D$ that defines the correlation length: $\lambda_c \sim 1/|D|^{1/\varphi}$. 

Furthermore, at $|\xi| \gg 1$, the parameter $C$ can be safely set to zero resulting in the scaling
\bea
w_{s,d} &=& |D|^{1/\varphi} q_{s,d}^{(\pm)} ( |D|^{1/\varphi} \lambda )\qquad (|\xi| \gg 1)\, ,  \label{q}\\
g_{s,d} &=&  |D|^{2/\varphi} p_{s,d}^{(\pm)} ( |D|^{1/\varphi} \lambda ) \qquad (|\xi| \gg 1) \, , \label{p}
\eea 
where $q_{s,d}^{(\pm)}(x)$ and $p_{s,d}^{(\pm)}(x)$ are certain scaling functions with the superscripts corresponding to the sign of $D$ (see Fig.~\ref{fig:q_scaling}). The ``$-$" functions are related to their ``$+$" counterparts by the requirement that at any finite $\lambda$ the quantities  $w_{s,d}$ and $g_{s,d}$ be analytic functions of $D$.
The asymptotic behavior of these scaling functions in the $x\to 0$ limit is dictated by requirement that Eqs.~(\ref{q})--(\ref{p}) match Eqs.~(\ref{sol1})--(\ref{sol4}) with $C=C'=D'=0$. Note that the scaling functions $q_{d}^{(+)}$ and $q_{d}^{(-)}$ describe the flow of $w_d$ shown in Fig.~\ref{fig:w_s_dep}.

\begin{figure}
    \centering
    \includegraphics[width=\linewidth]{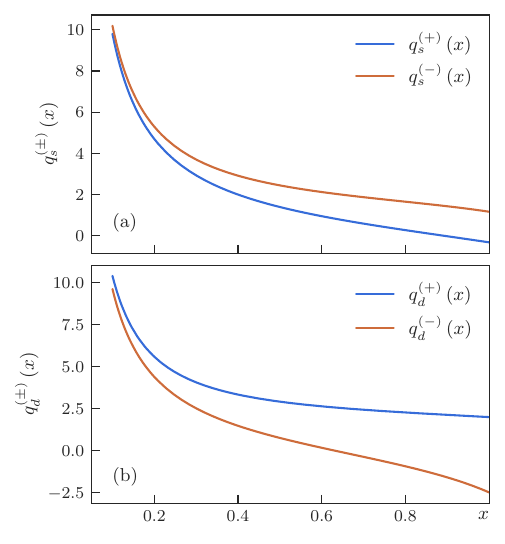}
    \caption{Scaling functions $q^{(\pm)}_{s,d}(x)$ defined in
Eq. (\ref{q}).}
    \label{fig:q_scaling}
\end{figure}

{\it Generalization}---The RG equations can readily be generalized to the case of an arbitrary number of components and interaction matrices.
Consider $N_G$ Goldstone modes with phases $\phi_\alpha$ with the generic action density
\begin{align}
    \mathcal{A}[\vec{\phi}] = \frac{1}{2} \nabla\vec{\phi}^T \hat{\Lambda} \nabla \vec{\phi} \, ,
\end{align}
where vector $\vec{\phi}$ has components $\phi_a$ and $\hat{\Lambda}$ is a real, symmetric, and positive definite matrix (i.e., we have gauged out all the redundant degrees of freedom with eigenvalue zero).

Any (single or composite) vortex in such a system will have integer topological charge in all $N_G$ components. So we can label them by a vector $\vec{M}$ with $M_{\alpha}$ being the (integer) charge in component $\alpha$.

Following similar calculations to those in 
Ref.~\cite{Golic2025}, we see that vortex-antivortex pairs of $\vec{M}$-type vortices will have energies proportional to $\vec{M}^T \hat{\Lambda} \vec{M} \equiv \Lambda_{\vec{M}}$. The general RG equations then read
\begin{align}
    \frac{\partial g_{\vec{M}}}{\partial\lambda} = g_{\vec{M}} (4 - 2 \pi \Lambda_{\vec{M}})\, , \\
    \frac{\partial\Lambda_{\alpha\beta}}{\partial\lambda} = - \sum_{\vec{M}} g_{\vec{M}} (\vec{M}\hat{\Lambda})_\alpha (\vec{M}\hat{\Lambda})_\beta ,
\end{align}
where $g_{\vec{M}}$ is the fugacity of vortex pair of type $\vec{M}$.
For the flow of $\Lambda_{\vec{M}}$ RG equation imply:
\begin{align}
    \frac{\partial\Lambda_{\vec{M}}}{\partial\lambda} = - \sum_{\vec{M}'} g_{\vec{M}'} \Big(\vec{M}^T\hat{\Lambda}\vec{M}'\Big)^2,
\end{align}
The transition away from superfluidity will then be driven by the proliferation of the (composite) vortex types with the smallest $\Lambda_{\vec{M}}$, with the relative fugacity of all other vortex types decaying to zero.
A question one can raise is how many topological excitations have the smallest $\Lambda_{\vec{M}}$ for different $\hat{\Lambda}$'s, and also what choice maximizes this number for different $N_G$.

As it turns out, this question can be mapped to a well-known problem for lattices.
Since $\hat{\Lambda}$ is real, symmetric, and positive definite, it is possible to find a matrix $\hat{A}$ with non-zero determinant such that $\hat{\Lambda} = \hat{A}^T\hat{A}$.
For any such matrix we can interpret $\hat{A}\vec{M}$ for integer vectors $\vec{M}$ as spanning an $N_G$-dimensional lattice, where each lattice point labeled by $\vec{M}$ has position $\hat{A}\vec{M}$ (the choice of $\hat{A}$ is not unique, although any matrices $\hat{A}, \hat{B}$ that satisfy $\hat{A}^T\hat{A} = \hat{B}^T\hat{B}$ are related by orthogonal transformations, and so represent fundamentally the same lattice).
Noting that $\Lambda_{\vec{M}} = \vec{M}^T \hat{A}^T\hat{A} \vec{M} = \|\hat{A} \vec{M}\|^2$ we see that the vortex pairs with minimal energy correspond to the sites that are the nearest neighbors of the origin in the lattice induced by $\hat{A}$.

Hence, for a given interaction matrix $\hat{\Lambda} = \hat{A}^T\hat{A}$ the maximal number of distinct elementary topological excitations (vortices) is equal to $N_v = \tau_{\hat{A}} / 2$ (we count vortex and antivortex as one species), where $\tau_{\hat{A}}$ is the kissing number of the lattice induced by $\hat{A}$. Thus, for $N_G > 1$, it is possible to find $\hat{\Lambda}$'s with substantially more topological defects than the number of Goldstone modes. An exact formula for the maximal kissing number for a lattice of arbitrary dimension is not known, although the number seems to grow rapidly with increasing dimension \cite{Conway1999}

As an example of a model with more than one extra elementary topological excitation, consider the following interaction matrix:
\begin{align}
\Lambda_{\alpha \alpha} = 2 \Lambda_s, \ \ \Lambda_{\alpha \neq \beta} = \Lambda_s \,.
\end{align}
This model has $N_v = N_G (N_G + 1) / 2$ elementary excitations given by $N_G$ single vortices and $N_G (N_G - 1) / 2$ composite vortices consisting of one vortex and one antivortex in different components: $\vec{M} = (0,...,1,0,...,-1,0,...)$.

\end{document}